\documentclass[12pt]{iopart}

\usepackage{iopams}  
\usepackage{graphicx}
\begin{document}

\title[Gap eigenmode of radially localised helicon waves]{Gap eigenmode of radially localised helicon waves in a periodic structure}

\author{L Chang$^1$, B N Breizman$^2$ and M J Hole$^1$}

\address{$^1$ Plasma Research Laboratory, Research School of Physics and Engineering, Australian National University, Canberra, ACT 0200, Australia}
\eads{\mailto{chang.lei@anu.edu.au}, \mailto{matthew.hole@anu.edu.au}}
\address{$^2$ Institute for Fusion Studies, The University of Texas, Austin, Texas 78712, USA}
\ead{breizman@mail.utexas.edu}

\begin{abstract}
An ElectroMagnetic Solver (EMS) [Chen \etal, Phys. Plasmas, 13, 123507 (2006)] is employed to model a spectral gap and a gap eigenmode in a periodic structure in the whistler frequency range. A Radially Localised Helicon (RLH) mode [Breizman and Arefiev, Phys. Rev. Lett, 84, 3863 (2000)] is considered. We demonstrate that the computed gap frequency and gap width agree well with a theoretical analysis, and find a discrete eigenmode inside the gap by introducing a defect to the system's periodicity. The axial wavelength of the gap eigenmode is close to twice the system's periodicity, which is consistent with Bragg's law. Such an eigenmode could be excited by energetic electrons, similar to the excitation of Toroidal Alfv\'{e}n Eigenmodes (TAE) by energetic ions in tokamaks.
\end{abstract}

\maketitle

\section{Introduction}\label{introduction}
It is a generic phenomenon that spectral gaps are formed when waves propagate in periodic media, and eigenmodes can exist with frequencies inside the spectral gaps if a defect is introduced to break the system's perfect translational symmetry, and create an effective potential well to localise these waves.\cite{Strutt:1887aa, Anderson:1958aa, Mott:1968aa, John:1987aa, Yablonovitch:1991aa, Figotin:1997aa} Fusion plasmas have a few periodicities that can produce spectral gaps: geodesic curvature of field lines,\cite{Chu:1992aa} elongation\cite{DIppolito:1980aa, Dewar:1974aa} or triangularity of flux surfaces,\cite{Betti:1992aa} helicity\cite{Nakajima:1992aa} and periodic mirroring in stellerators.\cite{Kolesnichenko:2001aa} Weakly damped eigenmodes which are readily destabilised by energetic ions often reside in these gaps, and they may degrade fast ion confinement.\cite{Duong:1993aa, White:1995aa} The most extensively studied gap eigenmode is the toroidcity-induced Alfv\'{e}n eigenmode (TAE),\cite{Cheng:1985aa} however, there are also numerous other modes with similar features.\cite{Wong:1999aa}

Zhang \etal\cite{Zhang:2008aa} observed a spectral gap in the shear Alfv\'{e}n wave continuum in experiments on LArge Plasma Device (LAPD) with a multiple magnetic mirror array, and obtained consistent results through two-dimensional numerical modelling using the finite difference code, ElectroMagnetic Solver (EMS).\cite{Chen:2006aa} Although eigenmodes inside this gap were not formed, a possible experimental implementation was proposed to detect them. The idea is to use the endplate of the machine as a ``defect" that breaks the axial periodicity or to vary current in one of the independently powered magnetic coils.

In the present study, we will use EMS to examine a gap eigenmode in a linear system with slightly broken axial periodicity. We will consider the Radially Localised Helicon (RLH) mode, whose radial structure has been described in \cite{Breizman:2000aa}. We will show that the computed gap frequency and gap width agree well with a theoretical analysis, and that there is a discrete eigenmode inside the gap. Such an eigenmode could be excited by energetic electrons, similarly to the excitation of Toroidal Alfv\'{e}n Eigenmodes (TAE) by energetic ions in tokamaks.

\section{Theoretical analysis}\label{analysis}
\subsection{Basic equations}
The spatial structure of RLH wave field and the corresponding dispersion relation can be found from (17) in \cite{Breizman:2000aa} that applies to a whistler-type linear wave in a cold plasma cylinder and has the form: 
\small
\begin{equation}
\frac{1}{r}\frac{\partial}{\partial r}r \frac{\partial E}{\partial r}-\frac{m^2}{r^2}E=-\frac{m}{k_z^2 r}\frac{\omega^2}{c^2}\frac{E\partial g/\partial r}{1+(m\partial g/\partial r)/k_z^2 r \eta},
\end{equation}
\normalsize
where $E=E_z-(k_z r/m)E_\varphi$ with $E_z$ and $E_\varphi$ the axial and azimuthal components of the wave electric field, respectively. Functions $g$ and $\eta$ are $g=\omega_{p}^2/\omega \omega_{c}$ and $\eta=-\omega_{p}^2/\omega^2$, respectively, where $\omega_{p}$ is the electron plasma frequency and $\omega_{c}$ the electron cyclotron frequency. 

In this section, we will limit our consideration to the case of sufficiently dense plasma in which $\omega_p/c\gg m/a$, where $a$ is the plasma radius. This assumption makes it allowable to set $E_z=0$ and ignore the term $(m\partial g/\partial r)/k_z^2 r \eta$ in the denominator of (1), which gives
\small
\begin{equation}
k_z^2\left(\frac{\partial}{\partial r}r\frac{\partial}{\partial r}r E_\varphi-m^2 E_\varphi\right)=-E_\varphi m \frac{\omega^2}{c^2}r\frac{\partial g}{\partial r}.
\end{equation}
\normalsize
We now generalise this equation to the case of slightly modulated ($z$-dependent) plasma equilibrium with a following separable form of $\omega_p^2/\omega_c$:
\small
\begin{equation}
\frac{\omega_p^2}{\omega_c}=\frac{\omega_{p0}^2(r)}{\omega_{c0}}[1-\epsilon(z)\cos q z]. 
\end{equation}
\normalsize
Here, $\epsilon\ll 1$ and $q$ are the modulation envelope and wavenumber respectively. The axial scale-length of the envelope $\epsilon(z)$ is assumed to be much greater than $1/q$. We recall that $k_z^2=-\partial^2/\partial z^2$ and transform (2) to 
\small
\begin{equation}
\frac{\partial^2}{\partial z^2}\left(\frac{\partial}{\partial r}r\frac{\partial}{\partial r}r E_\varphi-m^2 E_\varphi\right)=E_\varphi m \frac{\omega}{c^2}[1-\epsilon(z)\cos q z]r\frac{\partial}{\partial r}\left(\frac{\omega_{p0}^2}{\omega_{c0}}\right).
\end{equation}
\normalsize
The modulated equilibrium introduces resonant coupling between the modes with $k_z=q/2$ and $k_z=-q/2$, which suggests the following form for $E_\varphi$: 
\small
\begin{equation}
E_\varphi=A_+e^{iqz/2}+A_-e^{-iqz/2}, 
\end{equation}
\normalsize
where $A_+$ and $A_-$ are slow functions of $z$ compared to $\cos qz$. 

Let $\omega_0$ be an eigenfrequency of (4) for $\epsilon=0$ and $|k_z|=q/2$. It is then straightforward to separate spatial scales in (4) and obtain a set of coupled equations for $A_+$ and $A_-$:
\small
\begin{equation}
\begin{array}{l}
\hspace{-1cm}\vspace{0.2cm}-\frac{q^2}{4}\left(\frac{\partial}{\partial r}r\frac{\partial}{\partial r}r A_+-m^2 A_+\right)-A_+m\frac{\omega_0}{c^2}r\frac{\partial}{\partial r}\left(\frac{\omega_{p0}^2}{\omega_{c0}}\right)=\\
\hspace{-1cm}-i q\frac{\partial}{\partial z}(\frac{\partial}{\partial r}r \frac{\partial}{\partial r}r A_+-m^2 A_+)+A_+ m \frac{\omega-\omega_0}{c^2}r\frac{\partial}{\partial r}\left(\frac{\omega_{p0}^2}{\omega_{c0}}\right)-\frac{1}{2}A_-m \frac{\omega_0}{c^2}\epsilon(z)r\frac{\partial}{\partial r}\left(\frac{\omega_{p0}^2}{\omega_{c0}}\right),
\end{array}
\end{equation}
\normalsize
\small
\begin{equation}
\begin{array}{l}
\hspace{-1cm}\vspace{0.2cm}-\frac{q^2}{4}\left(\frac{\partial}{\partial r}r\frac{\partial}{\partial r}r A_--m^2 A_-\right)-A_- m\frac{\omega_0}{c^2}r\frac{\partial}{\partial r}\left(\frac{\omega_{p0}^2}{\omega_{c0}}\right)=\\
\hspace{-1cm}+i q\frac{\partial}{\partial z}\left(\frac{\partial}{\partial r}r \frac{\partial}{\partial r}r A_--m^2 A_-\right)+A_- m \frac{\omega-\omega_0}{c^2}r\frac{\partial}{\partial r}\left(\frac{\omega_{p0}^2}{\omega_{c0}}\right)-\frac{1}{2}A_+ m \frac{\omega_0}{c^2}\epsilon(z)r\frac{\partial}{\partial r}\left(\frac{\omega_{p0}^2}{\omega_{c0}}\right).
\end{array}
\end{equation}
\normalsize
We have neglected second axial derivatives of $A_+$ and $A_-$ and arranged (6) and (7) so that their right hand sides represent relatively small terms compared to the dominant terms on the left hand sides. With this ordering, we conclude that the radial dependencies of $A_+$ and $A_-$ need to be close to the eigenfunction $\Psi(r)$ of the lowest order ODE
\small
\begin{equation}
\frac{q^2}{4}\left(\frac{\partial}{\partial r}r\frac{\partial}{\partial r}r \Psi-m^2 \Psi\right)+\Psi m\frac{\omega_0}{c^2}r\frac{\partial}{\partial r}\left(\frac{\omega_{p0}^2}{\omega_{c0}}\right)=0.
\end{equation}
\normalsize
The differential operator in this equation is self-adjoint. As a result, multiplication of (6) and (7) by $r\Psi$ and integration over radius lead to
\small
\begin{equation}
\begin{array}{ll}
\hspace{-0.1cm}\vspace{0.2cm}0=&-i q\frac{\partial}{\partial z}\int dr r\Psi\left(\frac{\partial}{\partial r}r \frac{\partial}{\partial r}r A_+-m^2 A_+\right)+\int dr r\Psi A_+ m \frac{\omega-\omega_0}{c^2}r\frac{\partial}{\partial r}\left(\frac{\omega_{p0}^2}{\omega_{c0}}\right)\\
&-\frac{1}{2}\int dr r\Psi A_-m \frac{\omega_0}{c^2}\epsilon(z)r\frac{\partial}{\partial r}\left(\frac{\omega_{p0}^2}{\omega_{c0}}\right),
\end{array}
\end{equation}
\normalsize
\small
\begin{equation}
\begin{array}{ll}
\hspace{-0.1cm}\vspace{0.2cm}0=&i q\frac{\partial}{\partial z}\int dr r\Psi\left(\frac{\partial}{\partial r}r \frac{\partial}{\partial r}r A_--m^2 A_-\right)+\int dr r\Psi A_- m \frac{\omega-\omega_0}{c^2}r\frac{\partial}{\partial r}\left(\frac{\omega_{p0}^2}{\omega_{c0}}\right)\\
&-\frac{1}{2}\int dr r\Psi A_+ m \frac{\omega_0}{c^2}\epsilon(z)r\frac{\partial}{\partial r}\left(\frac{\omega_{p0}^2}{\omega_{c0}}\right).
\end{array}
\end{equation}
\normalsize
Having eliminated the lowest order terms, we now set
\small
\begin{equation}
A_+=F(z)\Psi(r),~A_-=G(z)\Psi(r)
\end{equation}
\normalsize
in (9) and (10) to obtain the following set of coupled equations for $F(z)$ and $G(z)$: 
\small
\begin{equation}
0=\frac{4 i}{q}\frac{\partial F}{\partial z}+F\frac{\omega-\omega_0}{\omega_0}-\frac{G}{2}\epsilon(z), 
\end{equation}
\normalsize
\small
\begin{equation}
0=-\frac{4 i}{q}\frac{\partial G}{\partial z}+G\frac{\omega-\omega_0}{\omega_0}-\frac{F}{2}\epsilon(z). 
\end{equation}
\normalsize

\subsection{Spectral gap and continuum}
In the case of $z$-independent $\epsilon$, (12) and (13) have exponential solutions
\small
\begin{equation}
F\propto G\propto e^{i\kappa z}
\end{equation}
\normalsize
with $\kappa$ the wave number. The corresponding dispersion relation has two roots: 
\small
\begin{equation}
\begin{array}{l}
\hspace{-0.1cm}\vspace{0.2cm}\omega_+(\kappa)=\omega_0[1+\sqrt{(4\kappa/q)^2+(\epsilon/2)^2}],\\
\hspace{-0.1cm}\omega_-(\kappa)=\omega_0[1-\sqrt{(4\kappa/q)^2+(\epsilon/2)^2}].
\end{array}
\end{equation}
\normalsize
Here, $\omega_+(\kappa)$ and $\omega_-(\kappa)$ are the continuum frequencies above and below the spectral gap, respectively. The upper and lower tips of the spectral gap correspond to $\kappa=0$:
\small
\begin{equation}
\begin{array}{l}
\hspace{-0.1cm}\vspace{0.2cm}\omega_+(0)=\omega_0\left(1+\frac{\epsilon}{2}\right),\\
\hspace{-0.1cm}\omega_-(0)=\omega_0\left(1-\frac{\epsilon}{2}\right).
\end{array}
\end{equation}
\normalsize
The normalised width of the gap, $\Delta\omega\equiv[\omega_+(0)-\omega_-(0)]/\omega_0$, is therefore equal to the modulation amplitude $\epsilon$. The central frequency of the gap $\omega_0$ needs to be found from (8) as the RLH eigenfrequency for an axially uniform plasma cylinder and it can be written as
\small
\begin{equation}
\omega_0=\Gamma \frac{\omega_{c0}(0)c^2 q^2}{4\omega_{p0}(0)^2},
\end{equation} 
\normalsize
where $\omega_{c0}(0)$ and $\omega_{p0}(0)$ are the on-axis values and the numerical form-factor $\Gamma$ is determined by the plasma radial profile. 

\subsection{Wall-localised eigenmodes}\label{wall}
A discrete-spectrum eigenmode can be created inside the spectral gap if the system's periodicity is broken. We illustrate that by considering an ideally conducting endplate located at $z=z_0$, so that the plasma now occupies only a half cylinder to the right of the endplate ($z>z_0$). The boundary condition at the endplate is
\small
\begin{equation}
E_\varphi(r; z_0)=0
\end{equation}
\normalsize
or, equivalently, 
\small
\begin{equation}
F(z_0)e^{i q z_0/2}+G(z_0)e^{-i q z_0/2}=0
\end{equation}
\normalsize
(see (5) and (11)). The electric field of the discrete-spectrum mode must also vanish at $z\rightarrow\infty$, i. e., 
\small
\begin{equation}
F(\infty)=G(\infty)=0. 
\end{equation}
\normalsize
Equations (12) and (13) (with $\epsilon=const$) admit an exponential solution
\small
\begin{equation}
F\propto G\propto e^{-\lambda z}
\end{equation}
\normalsize
with a positive value of $\lambda$ that satisfies boundary conditions in (19) and (20). 

Indeed, the exponential ansatz, (21), reduces (12), (13) and (19) to an algebraic set
\small
\begin{equation}
0=-\frac{4 i\lambda}{q}F+F\frac{\omega-\omega_0}{\omega_0}-\frac{G}{2}\epsilon, 
\end{equation}
\normalsize
\small
\begin{equation}
0=\frac{4 i\lambda}{q}G+G\frac{\omega-\omega_0}{\omega_0}-\frac{F}{2}\epsilon,
\end{equation}
\begin{equation}
F+G e^{-i q z_0}=0.
\end{equation}
\normalsize
The solvability conditions for this set determine $\omega$ and $\lambda$ as functions of $z_0$ as follows: 
\small
\begin{equation}
\frac{\omega-\omega_0}{\omega_0}=-\frac{\epsilon}{2}\cos q z_0,
\end{equation}
\normalsize
\small
\begin{equation}
\lambda=\frac{\epsilon}{8}q\sin q z_0. 
\end{equation}
\normalsize
We observe that $\lambda$ is positive for $0<q z_0<\pi$, and the wave field therefore vanishes at infinity in this case to represent a discrete eigenmode whose frequency is automatically inside the gap. In particular, the eigenfrequency is exactly at the gap centre when $q z_0=\pi/2$. We further note that this wall-localised solution can also be viewed as an odd-parity eigenmode ($E_\varphi(z)=-E_\varphi(2 z_0-z)$) in the entire periodic cylinder with a defect at $z=z_0$. An axial profile of $\omega_{p}^2/\omega_c$ with such defect is shown in figure 1(a). 
\begin{figure}[ht]
\begin{center}$
\begin{array}{ll}
\hspace{2.3cm}(a)&(b)\\
\hspace{2.3cm}\includegraphics[width=0.41\textwidth,angle=0]{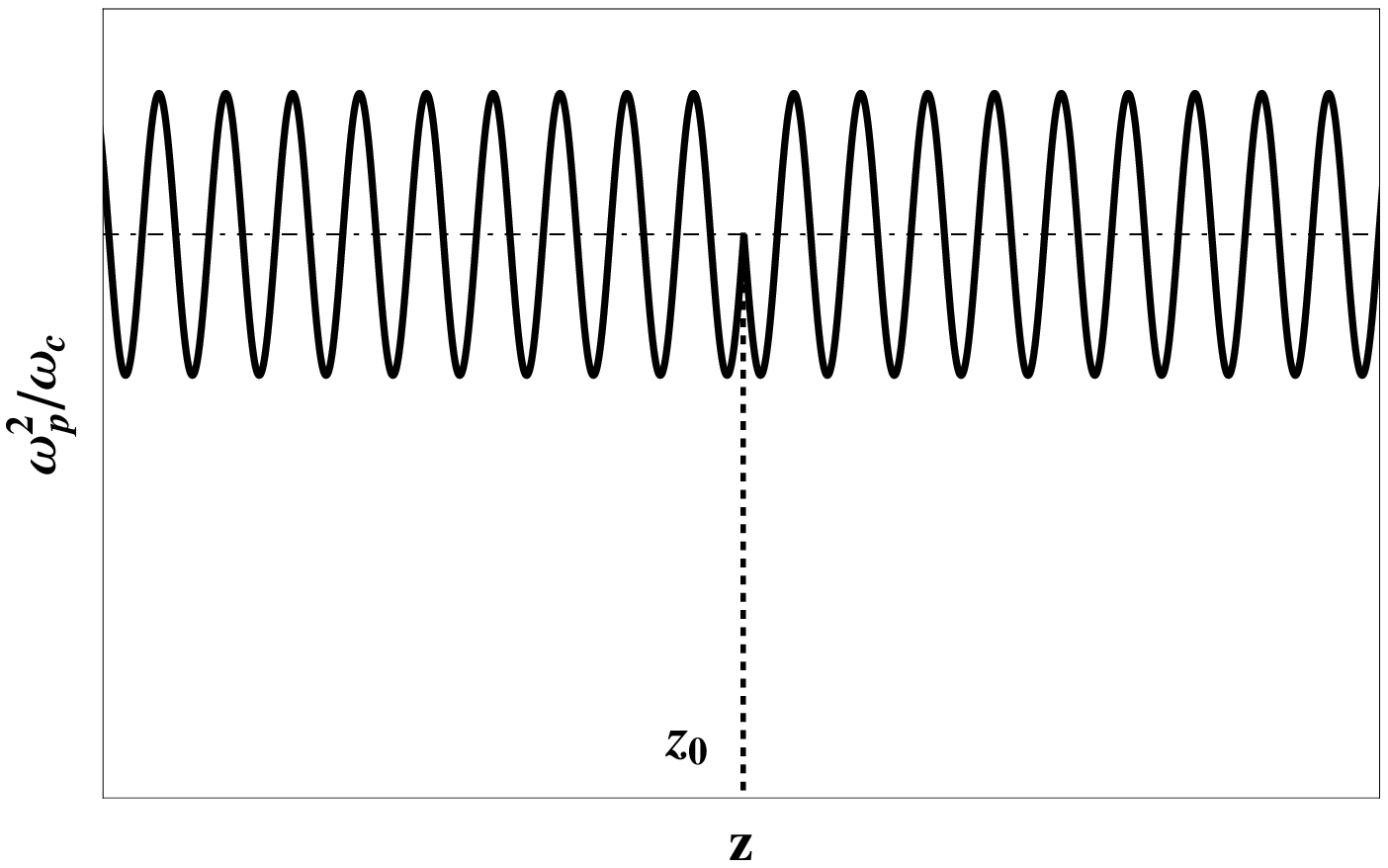}&\hspace{-0.07cm}\includegraphics[width=0.41\textwidth,height=0.259\textwidth,angle=0]{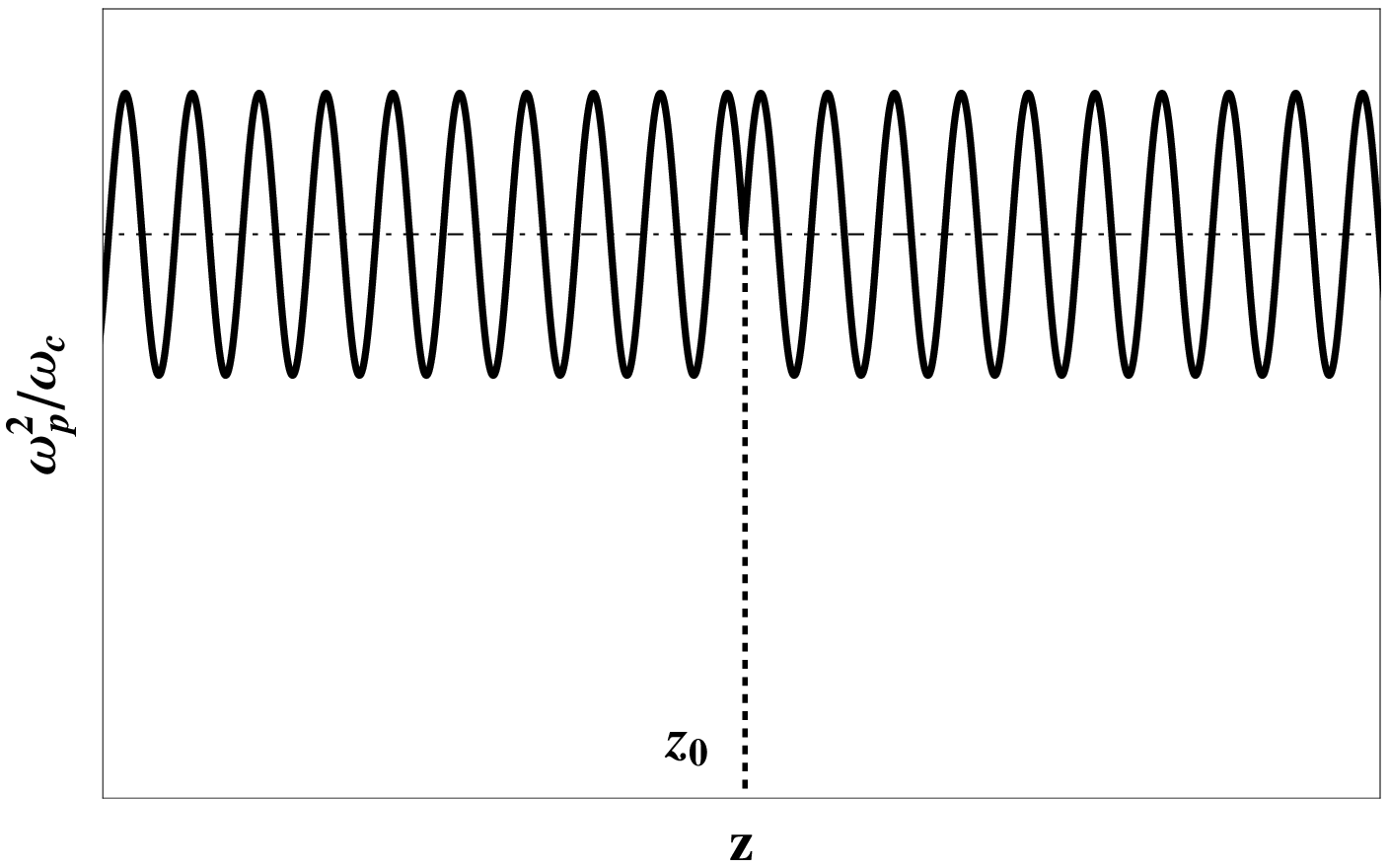}
\end{array}$
\end{center}
\caption{Axial profiles of $\omega_p^2/\omega_c$ with defects: (a) defect location at $\sin q z_0=1$, (b) defect location at $\sin q z_0=-1$.}
\label{defect}
\end{figure}

\subsection{Even-parity eigenmode}\label{even}
In contrast with figure 1(a), figure 1(b) illustrates a defect with $\pi<q z_0<2\pi$ that does not produce an odd-parity mode. Instead, an even-parity mode exists in this case. This mode is still described by (12) and (13), but the boundary condition at $z=z_0$ now changes from (18) to
\small
\begin{equation}
\left[\frac{\partial E_\varphi (r; z)}{\partial z}\right]_{z=z_0}=0,
\end{equation}
\normalsize
or, equivalently,
\small
\begin{equation}
F-Ge^{-i q z_0}=0.
\end{equation}
\normalsize
The solvability conditions for (22), (23) and (28) now give
\small
\begin{equation}
\frac{\omega-\omega_0}{\omega_0}=\frac{\epsilon}{2}\cos q z_0,
\end{equation}
\normalsize
\small
\begin{equation}
\lambda=-\frac{\epsilon}{8}q\sin q z_0,
\end{equation}
\normalsize
and we observe that $\lambda$ is positive (the wave field vanishes at infinity) when $\sin q z_0<0$, i. e. this even-parity eigenmode indeed requires $\pi<q z_0<2\pi$ for its existence. 

\subsection{Schr\"{o}dinger equation for gap modes}
Equations (12) and (13) can be straightforwardly transformed into two independent second-order equations for $F-G$ and $F+G$: 
\small
\begin{equation}
\frac{16}{q^2}\frac{\partial}{\partial z}\frac{1}{\left(\frac{\omega-\omega_0}{\omega_0}-\frac{\epsilon}{2}\right)}\frac{\partial (F-G)}{\partial z}+(F-G)\left(\frac{\omega-\omega_0}{\omega_0}+\frac{\epsilon}{2}\right)=0,
\end{equation}
\begin{equation}
\frac{16}{q^2}\frac{\partial}{\partial z}\frac{1}{\left(\frac{\omega-\omega_0}{\omega_0}+\frac{\epsilon}{2}\right)}\frac{\partial (F+G)}{\partial z}+(F+G)\left(\frac{\omega-\omega_0}{\omega_0}-\frac{\epsilon}{2}\right)=0.
\end{equation}
\normalsize
Both of them can be further reduced to a time-independent Schr\"{o}dinger equation when $\epsilon$ is nearly constant. We assume $\epsilon=\langle\epsilon\rangle+u(z)$ with $u(z)\ll\langle\epsilon\rangle$ and $u(\pm\infty)=0$. The discrete-spectrum modes in this case are very close to the tips of the gap. For the lower tip, we have
\small
\begin{equation}
\frac{\omega-\omega_0}{\omega_0}=-\frac{\langle\epsilon\rangle}{2}+\delta_-
\end{equation}
\normalsize
with $|\delta|\ll\langle\epsilon\rangle$. We can then neglect $u(z)$ and $\delta_-$ in the derivative term of (31) and get
\small
\begin{equation}
-\frac{\partial^2 (F-G)}{\partial z^2}+\frac{q^2\langle\epsilon\rangle}{16}(F-G)\left[\delta_-+\frac{u(z)}{2}\right]=0. 
\end{equation}
\normalsize
Similar procedure for the upper tip gives
\small
\begin{equation}
-\frac{\partial^2 (F+G)}{\partial z^2}+\frac{q^2\langle\epsilon\rangle}{16}(F+G)\left[-\delta_++\frac{u(z)}{2}\right]=0. 
\end{equation}
\normalsize
We note that any negative function $u(z)$ acts as a potential well that supports spatially localised eigenmodes of (34) and (35) with $\delta_->0$ and $\delta_+<0$, respectively. The eigenfrequencies of these modes belong to the spectral gap in the continuum. 

\section{Numerical implementation} 
An ElectroMagnetic Solver (EMS)\cite{Chen:2006aa} based on Maxwell's equations and a cold plasma dielectric tensor is employed to study the RLH spectral gap and gap eigenmode inside. The Maxwell's equations are expressed in the frequency domain:
\small
\begin{equation}
\bigtriangledown\times\mathbf{E}=i \omega\mathbf{B}, 
\end{equation}
\begin{equation}
\frac{1}{\mu_0}\bigtriangledown\times\mathbf{B}=-i \omega \mathbf{D}+\mathbf{j_a},
\end{equation}
\normalsize
where $\mathbf{E}$ and $\mathbf{B}$ are the wave electric and magnetic fields, respectively, $\mathbf{D}$ is the electric displacement vector, $\mathbf{j_a}$ is the antenna current, and $\omega$ is the antenna driving frequency. These equations are Fourier transformed with respect to the azimuthal angle and then solved (for an azimuthal mode number $m$) by a finite difference scheme on a 2D domain ($r$; $z$). The computational domain is shown in figure 2.
\begin{figure}[ht]
\begin{center}
\hspace{0.6cm}\includegraphics[width=0.7\textwidth,angle=0]{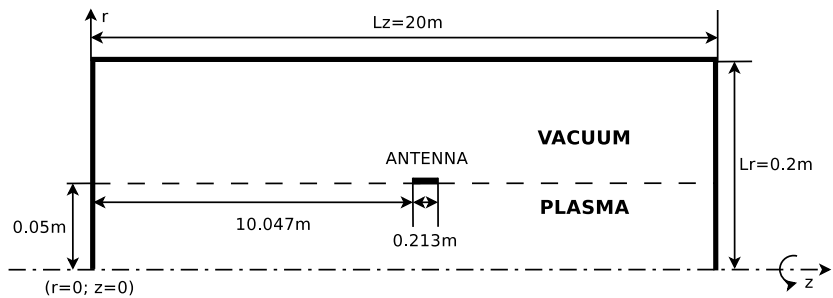}
\end{center}
\caption{Computational domain. The solid bar denotes a half-turn helical antenna. The dot-dashed line is the machine and coordinate system axis ($r=0$). The coordinate system ($r$; $\varphi$; $z$) is right-handed with an azimuthal angle $\varphi$.}
\label{domain}
\end{figure}
The quantities $\mathbf{D}$ and $\mathbf{E}$ are linked via a dielectric tensor,\cite{Ginzburg:1964aa} as follows:
\small
\begin{equation}
\mathbf{D}=\varepsilon_0[\varepsilon\mathbf{E}+ig(\mathbf{E}\times\mathbf{b})+(\eta-\varepsilon)(\mathbf{E}\cdot\mathbf{b})\mathbf{b}],
\end{equation}
\normalsize
where $\mathbf{b}\equiv\mathbf{B_0}/B_0$ is the unit vector along the static magnetic field and 
\small
\begin{equation}
\begin{array}{l}
\vspace{0.15cm}\varepsilon=1-\sum\limits_{\alpha}\frac{\omega+i\nu_\alpha}{\omega}\frac{\omega^2_{p\alpha}}{(\omega+i\nu_\alpha)^2-\omega^2_{c\alpha}},\\
\vspace{0.15cm}g=-\sum\limits_{\alpha}\frac{\omega_{c\alpha}}{\omega}\frac{\omega^2_{p\alpha}}{(\omega+i\nu_\alpha)^2-\omega^2_{c\alpha}},\\
\eta=1-\sum\limits_{\alpha}\frac{\omega^2_{p\alpha}}{\omega(\omega+i\nu_\alpha)}.
\end{array}
\end{equation}
\normalsize
The subscript $\alpha$ labels particle species (electrons and ions); $\omega_{p\alpha}\equiv\sqrt{n_\alpha q_\alpha^2/\varepsilon_0 m_\alpha}$ is the plasma frequency, $\omega_{c\alpha}\equiv q_\alpha B_0/m_\alpha$ is the cyclotron frequency, and $\nu_\alpha$ a phenomenological collision frequency for each species. The static magnetic field is assumed to be axisymmetric with $B_{0r}~\ll~B_{0z}$ and $B_{0\varphi}=0$. It is therefore appropriate to use a near axis expansion for the field, namely $B_{0z}$ is only dependent on $z$ and 
\small
\begin{equation}
B_{0r}(r,~z)=-\frac{1}{2}r\frac{\partial B_{0z} (z)}{\partial z}. 
\end{equation}
\normalsize
A helical antenna is employed to excite an $m=1$ mode in the plasma. The enclosing chamber is assumed to be ideally conducting so that the tangential components of $\mathbf{E}$ vanish at the chamber walls, i. e. 
\small
\begin{equation}
\begin{array}{l}
\vspace{0.2cm}E_\varphi(L_r; z)=E_z(L_r; z)=0, \\
\vspace{0.2cm}E_r(r; 0)=E_\varphi(r; 0)=0, \\
E_r(r; L_z)=E_\varphi(r; L_z)=0,  
\end{array}
\end{equation}
\normalsize
with $L_r$ and $L_z$ the radius and the length of the chamber, respectively. Further, all field components must be regular on axis.

\section{Numerical results and discussion}
\subsection{RLH mode in a straight cylinder}\label{uniform}
We first use EMS to calculate the radial structure of the RLH mode in a straight cylinder with a uniform static magnetic field and recover the mode dispersion relation. We consider a single-ionised argon plasma with a radial density profile shown in figure~3(a).
\begin{figure}[ht]
\begin{center}$
\begin{array}{ll}
\hspace{2.3cm}(a)&(b)\\
\hspace{2.3cm}\includegraphics[width=0.4\textwidth,angle=0]{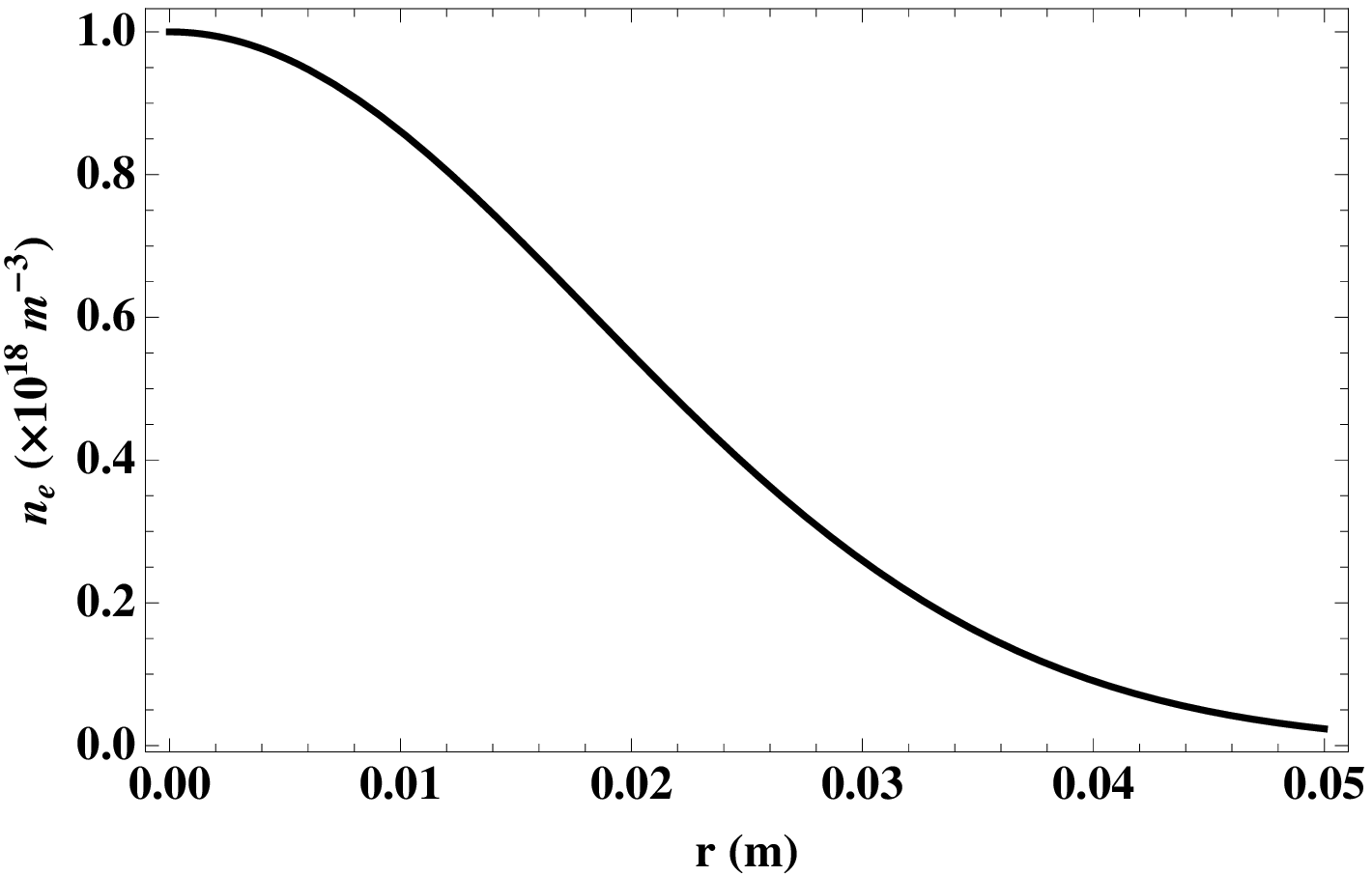}&\hspace{0.2 cm}\includegraphics[width=0.41\textwidth,height=0.2585\textwidth,angle=0]{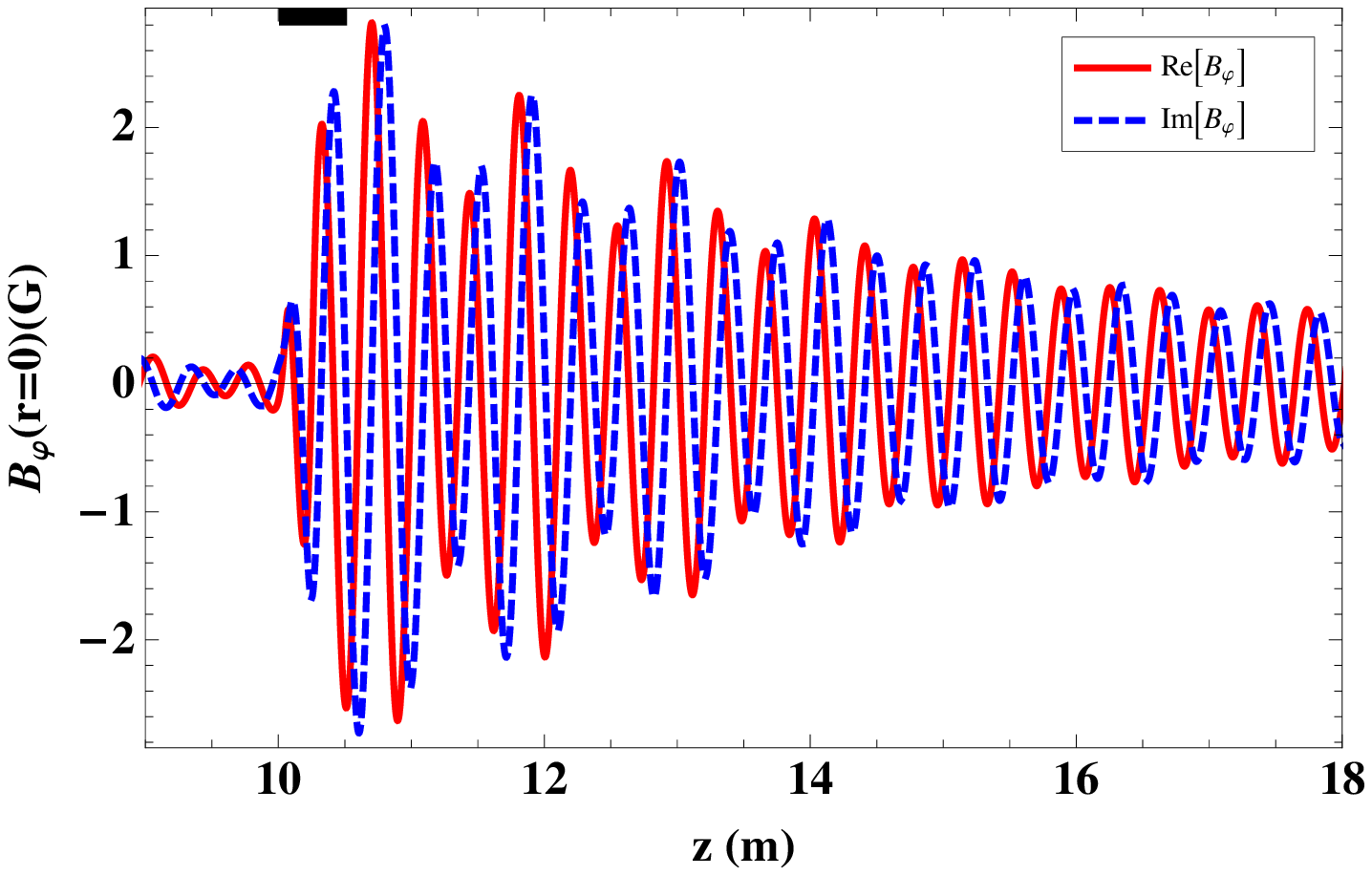}\\
\hspace{2.3cm}(c)&(d)\\
\hspace{2.3cm}\includegraphics[width=0.4\textwidth,angle=0]{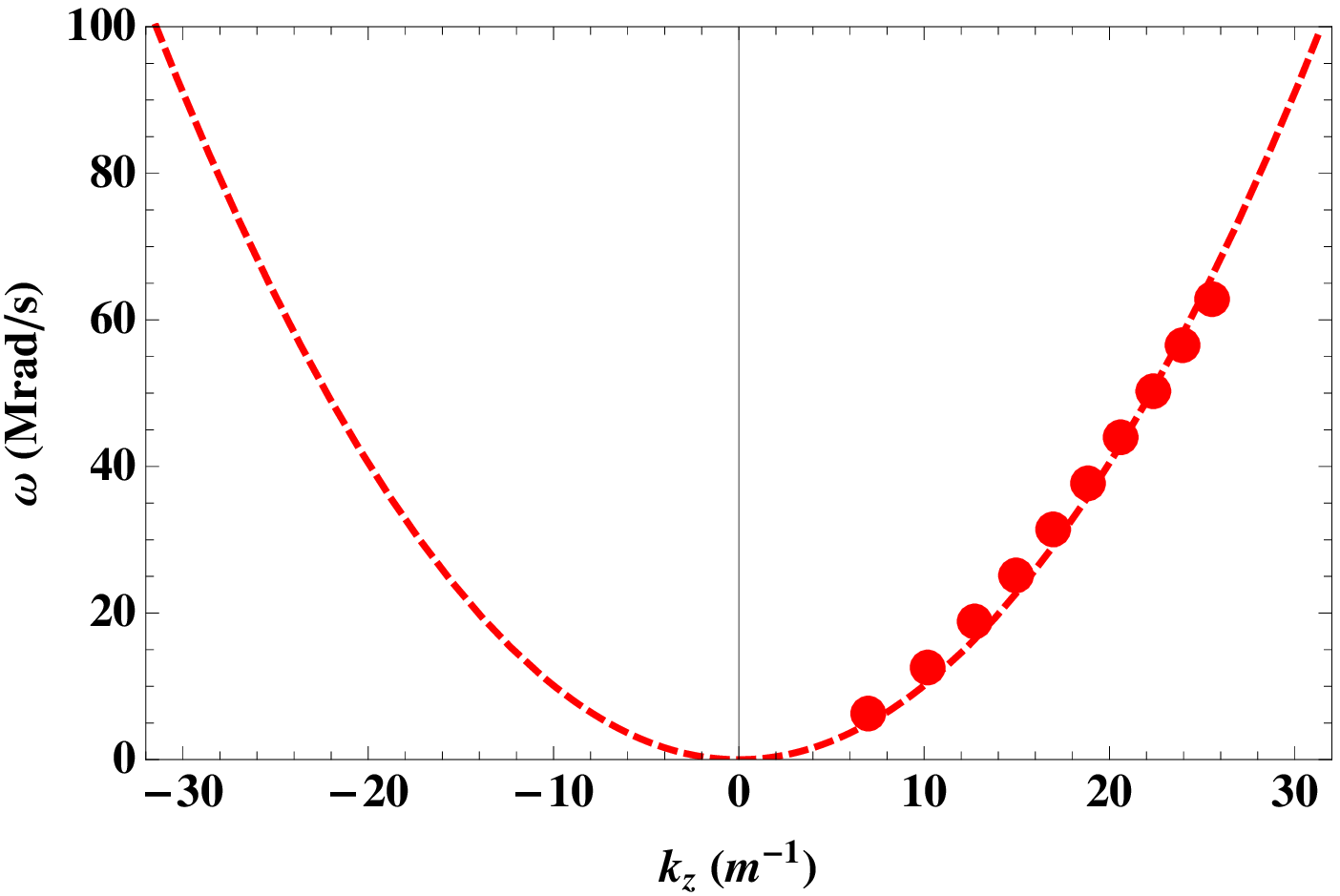}&\hspace{0.04cm}\includegraphics[width=0.41\textwidth,height=0.264\textwidth,angle=0]{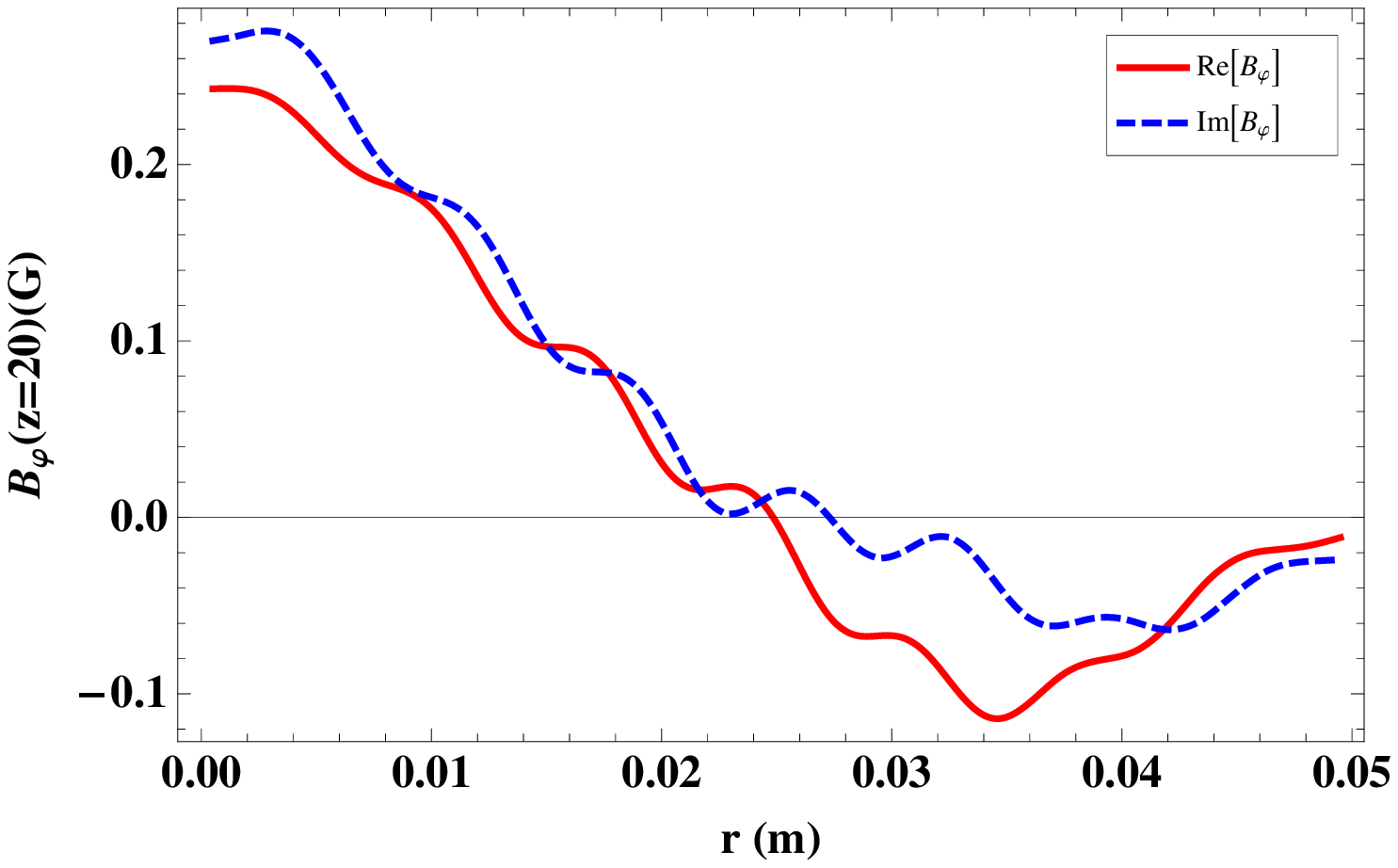}
\end{array}$
\end{center}
\caption{Plasma density profile and RLH wave field in a straight cylinder: (a) radial profile of unperturbed plasma density, (b) azimuthal magnetic field of the $m=1$ mode for an illustrative frequency $\omega=31.4$~Mrad/s (solid bar shows antenna location, and the endplates are located at $z=0$~m and $z=30$~m respectively), (c) computed dispersion relation for the $m=1$ mode (dots) and analytical scaling from (17) with $q/2=k_z$ and $\Gamma=2.03$ (dashed curve), (d) radial profile of the $m=1$ mode at  $z = 20$~m for $\omega=31.4$~Mrad/s. }
\label{coeff}
\end{figure}
The surrounding vacuum chamber and the position of the rf-antenna are shown in figure~2. The static magnetic field in this simulation is $B_0=0.01$~T and the electron-ion collision frequency on axis is $\nu_{ei}|_{r=0}=4\times10^6~\rm{s}^{-1}$. We run EMS for various frequencies of the antenna current, select the $m=1$ azimuthal component of the plasma response and calculate the dominant axial wavenumber, $k_z$, of this component via Fourier decomposition. We observe an apparently preferred axial propagation direction for the $m=1$ wave generated by the helical antenna. The collision frequency is sufficient to damp the radiated waves significantly towards the endplates but the dominant wavelength is still seen clearly in the plot of the excited field (figure 3(b)). The resulting relation between $k_z$ and $\omega$ is shown in figure 3(c). We note that $k_z$ scales as $\sqrt{\omega}$ with frequency, in agreement with analytical scaling obtained in \cite{Breizman:2000aa}. A least squares fit to the dispersion relation given by (17) with $q/2=k_z$ shows that the form factor $\Gamma$ equals $2.03$ for the selected density profile. The radial profile of this $m=1$ mode is presented in figure 3(d). 

\subsection{Spectral gap in a periodic structure}\label{gap}
In order to demonstrate the spectral gap, we run EMS for the same conditions as in section \ref{uniform} except that the static magnetic field is weakly modulated along $z$ and has the form $B_{0z} = B_0(1+0.25\cos qz)$ with $B_0=0.01$~T and $q=40~\rm{m}^{-1}$. The radial component of the static field is determined by (40). We scan the antenna frequency (for a fixed amplitude of the antenna current) and analyse wave propagation from the antenna. Figure 4 illustrates the results. Figure 4(a) is a plot of the on-axis signal at a $4.8$~m distance from the antenna, and shows an interval of strong suppression (from $36$~Mrad/s to $44$~Mrad/s), which represents the anticipated spectral gap. The location and width of this gap agree well with analytic estimates based on (16) and (17). Figure 4(b) presents spatial distributions of wave energy for frequencies that are inside and outside of the spectral gap, from which we also see that wave propagation is evanescent inside the gap.
\begin{figure}[ht]
\begin{center}$
\vspace{-0.3cm}
\begin{array}{l}
\hspace{2.3cm}(a)\\
\vspace{-0.3cm}\hspace{2.3cm}\includegraphics[width=0.655\textwidth,angle=0]{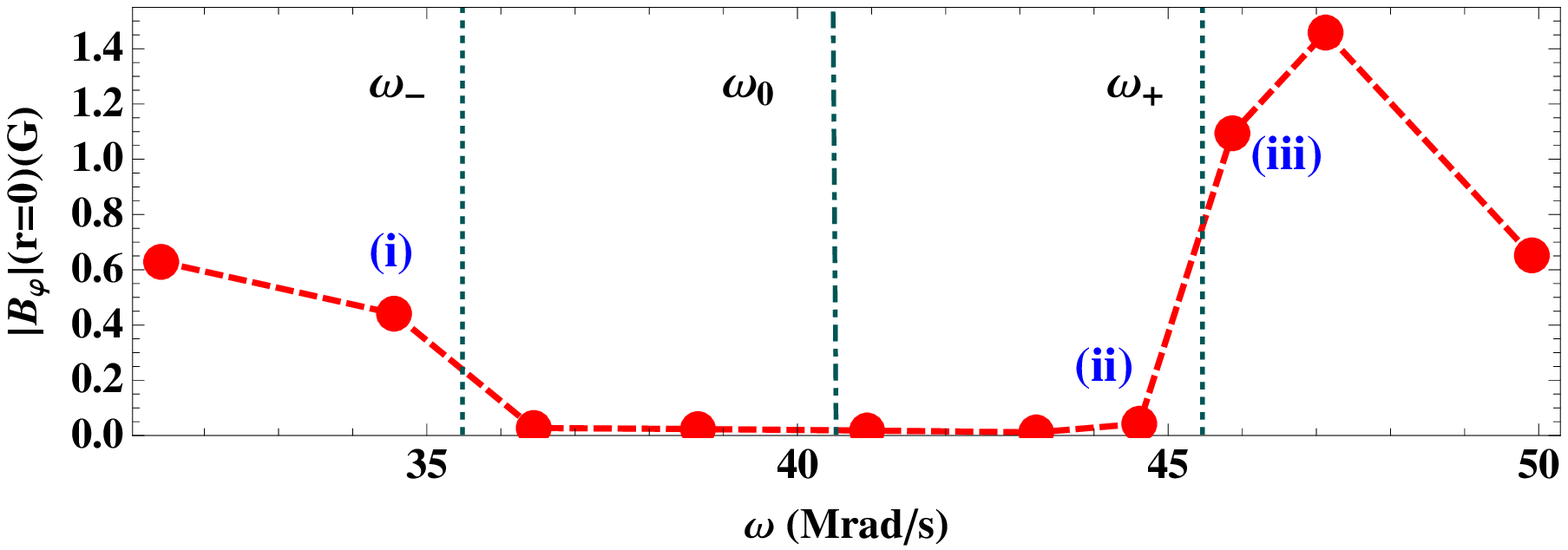}\\
\hspace{2.3cm}(b)\\
\vspace{-0.2cm}\hspace{2.2cm}\includegraphics[width=0.7\textwidth,angle=0]{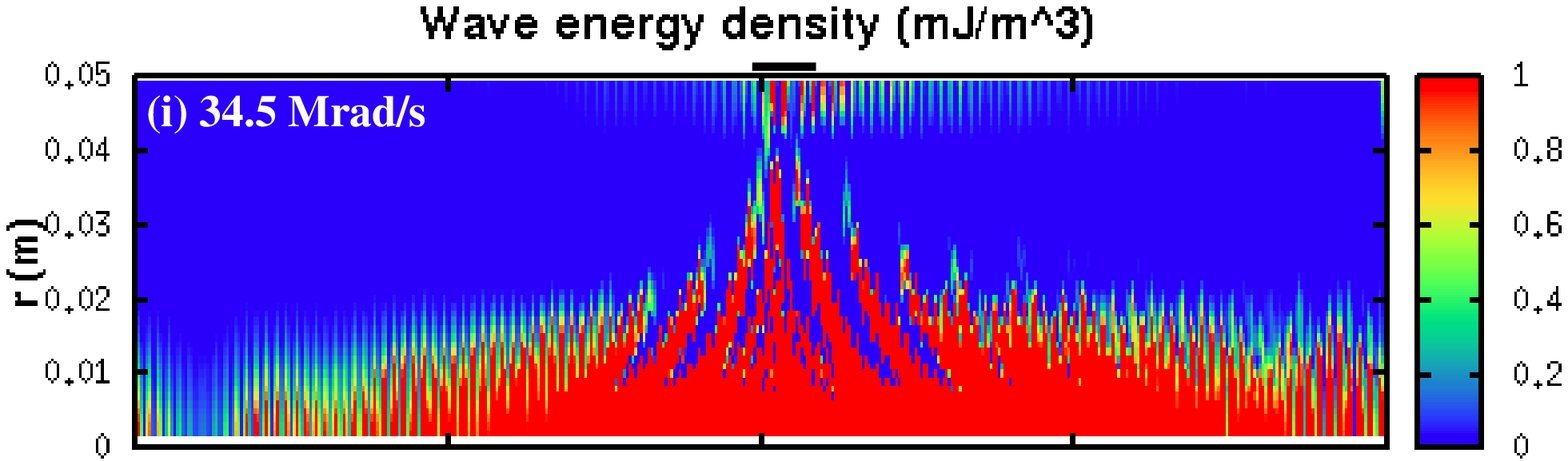}\\
\vspace{-0.2cm}\hspace{2.2cm}\includegraphics[width=0.7\textwidth,angle=0]{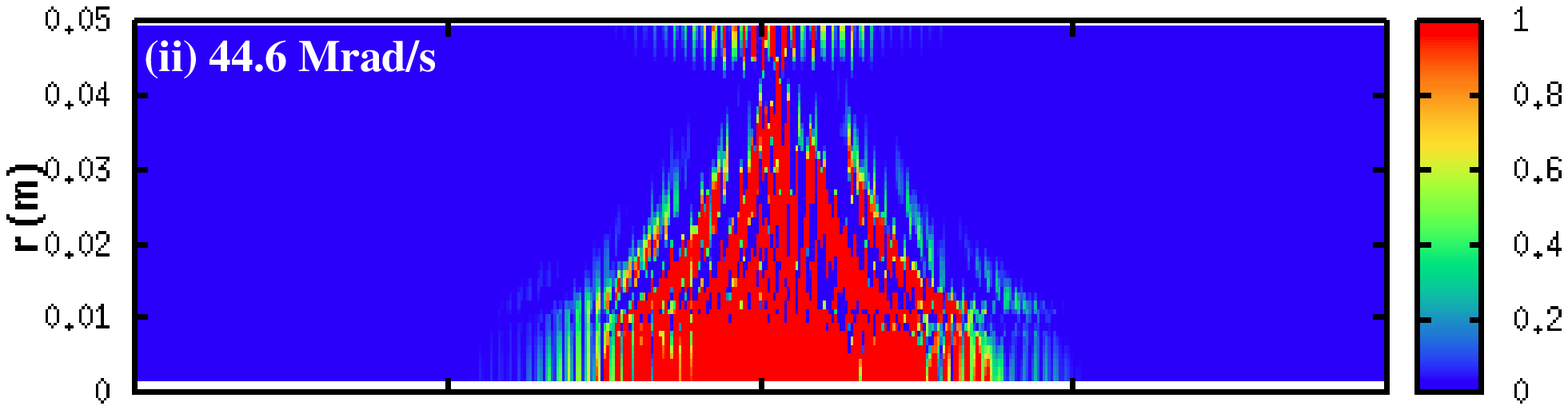}\\
\hspace{2.2cm}\includegraphics[width=0.7\textwidth,angle=0]{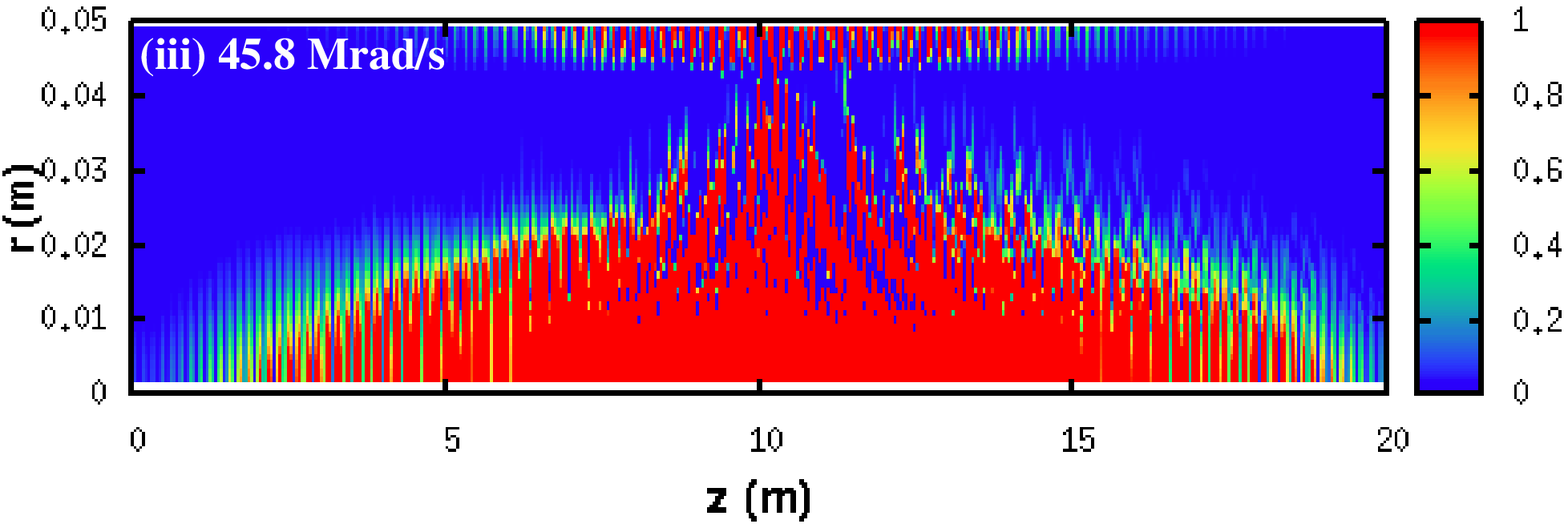}
\end{array}$
\end{center}
\caption{Identification of the spectral gap in simulations: (a) on-axis wave field at $z =15$~m as a function of driving frequency; vertical lines show analytical predictions for the gap centre $\omega_0$ and the upper and lower tips of the continuum $\omega_+$ and $\omega_-$, (b) spatial distributions of wave energy for antenna frequencies inside and outside the spectral gap (solid bar denotes the antenna location).}
\label{spectrum}
\end{figure}

\subsection{Gap modes}\label{gapmode}
To form an eigenmode inside the spectral gap shown in figure 4, we introduce a defect in the otherwise periodic static magnetic field. Equations (25) and (29) suggest that this defect should be located at $\cos q z_0= 0$, in order for the eigenmode frequency to be at the gap centre. The mode is then expected to have the shortest possible width (see (26) and (30)). Moreover, section \ref{wall} and section \ref{even} indicate that the gap eigenmode can have either odd parity or even parity, depending on the defect profile. We will confirm this analysis by simulation results shown in figure 5 and figure 6.
\begin{figure}[ht]
\begin{center}$
\begin{array}{ll}
\hspace{2.3cm}(a)&(b)\\
\hspace{2.32cm}\includegraphics[width=0.425\textwidth,angle=0]{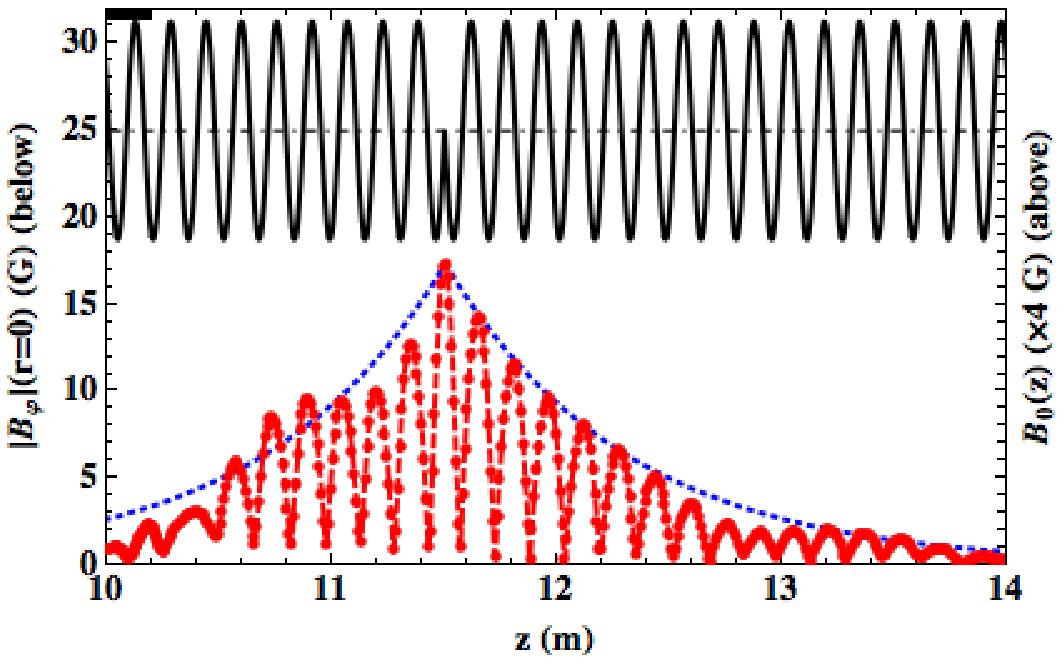}&\hspace{-0.3cm}\includegraphics[width=0.416\textwidth,height=0.262\textwidth,angle=0]{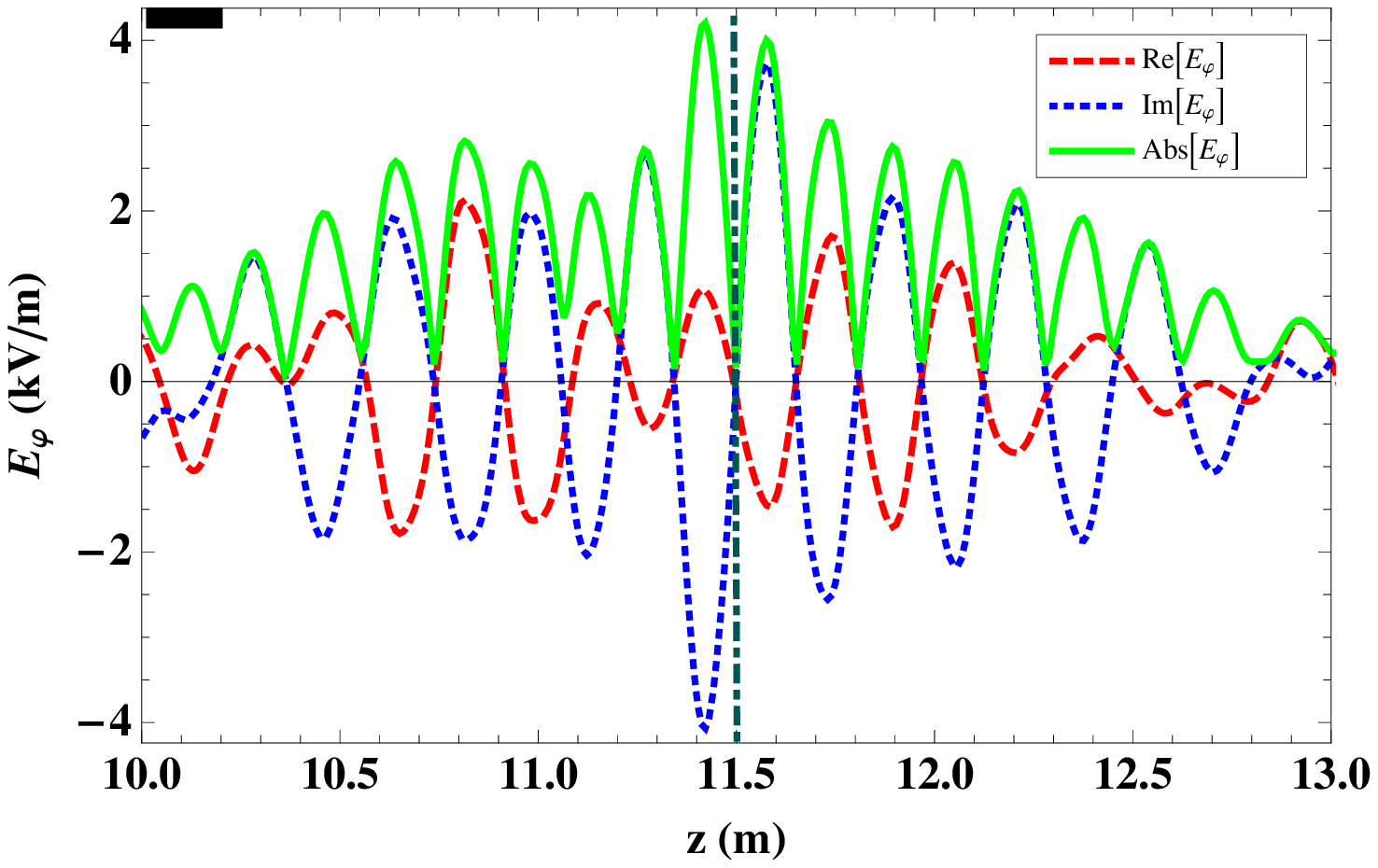}\\
\hspace{2.3cm}(c)&(d)\\
\hspace{2.4cm}\includegraphics[width=0.4\textwidth,height=0.27\textwidth,angle=0]{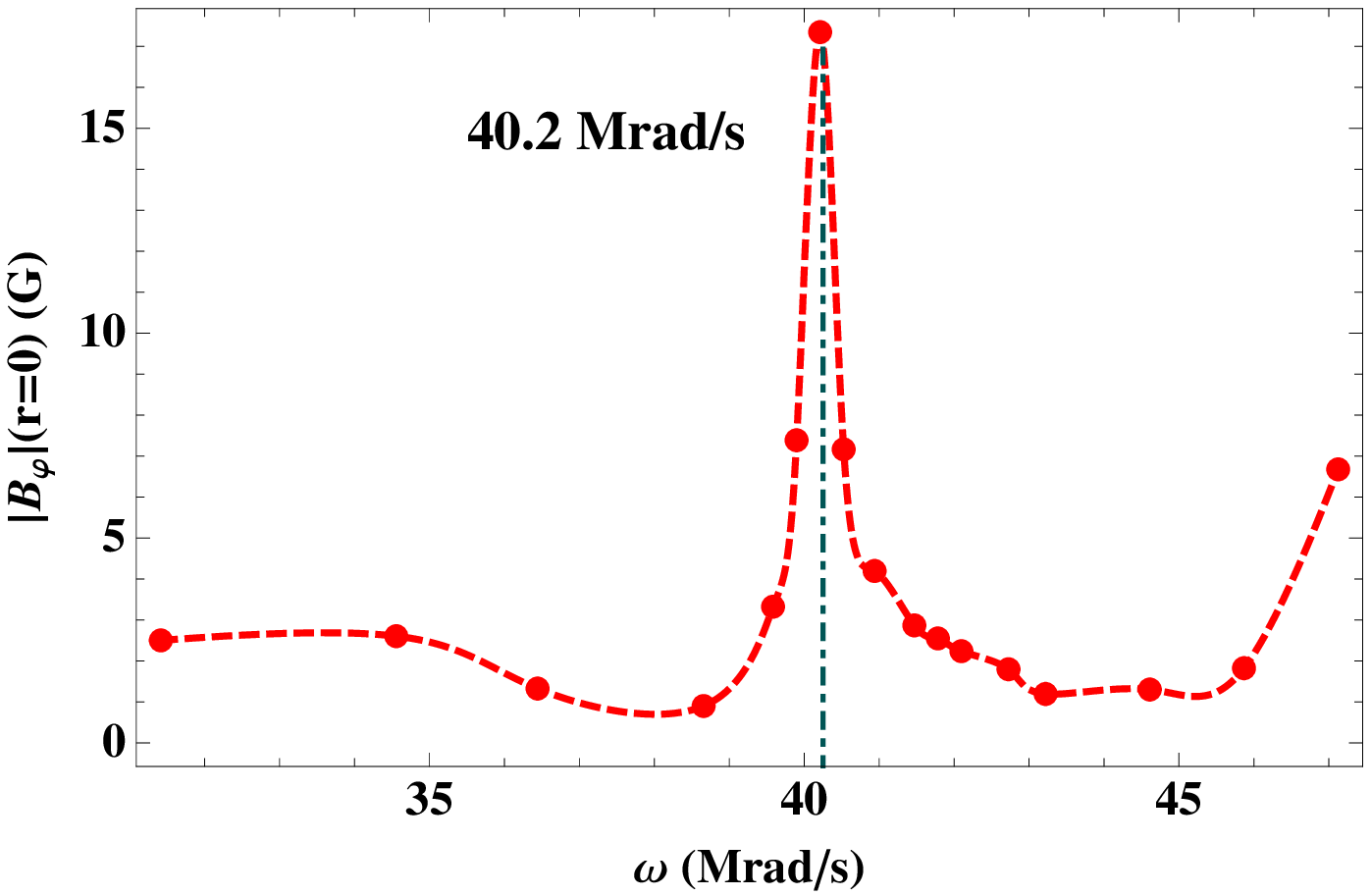}&\hspace{-0.5cm}\includegraphics[width=0.42\textwidth,angle=0]{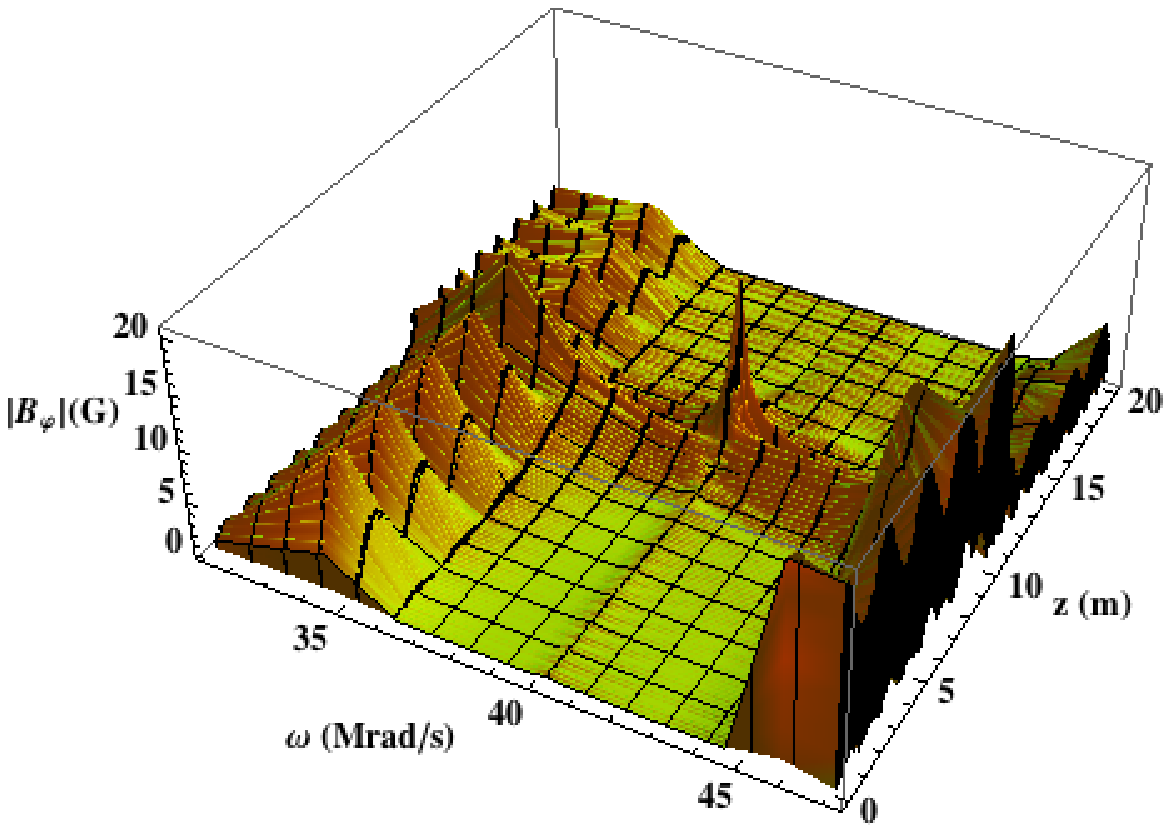}\\
\end{array}$
\end{center}
\caption{Odd-parity gap eigenmode: (a) longitudinal profiles of the static magnetic field (solid line) and the rf  magnetic field on axis for $\omega=40.2$~Mrad/s (dots), together with the theoretically calculated envelope (dotted line), (b) longitudinal profile of $E_\varphi$ (on-axis) at $\omega=40.2$~Mrad/s (vertical dot-dashed line marks the defect location, and solid horizontal bar marks the antenna), (c) resonance in the dependence of the on-axis amplitude of the rf magnetic field on driving frequency at the location of the defect, (d) 3D plot of the on-axis wave field strength as a function of $z$ and $\omega$. }
\label{oddmode}
\end{figure}

Figures 5(a)-(d) illustrate the odd-parity gap eigenmode associated with the defect shown in figure 1(a). In order to minimize the role of collisional dissipation and thereby obtain a sharper resonant peak in figure 5(c), we have reduced the collision frequency to $0.1\nu_{ei}$ in figure 5 simulation results. Figure 5(a) shows that the eigenmode is a standing wave localised around the defect. Its exponential envelope with a decay length of $1.25$~m is consistent with analytical expectation from (26). The evanescent field of the mode is only weakly coupled to the distant antenna, but the resonance with the antenna frequency still allows the mode to be excited easily. The internal scale of this mode is close to twice the system's periodicity, consistent with Bragg's law. Figure 5(b) shows the on-axis profile of the mode electric field, implying an odd function of $E_\varphi$ at $z_0$. The corresponding magnetic field $B_\varphi$ is, instead, an even function. Figure 5(d) presents a full view of the plasma response in ($z$; $\omega$) space with a clear eigenmode peak inside the spectral gap.

We have also performed similar calculations for the defect shown in figure 1(b). This defect produces an even-parity mode seen in figures 6(a)-(d). Except for different symmetries, the features of the even and odd modes are  apparently similar.
\begin{figure}[ht]
\begin{center}$
\begin{array}{ll}
\hspace{2.3cm}(a)&(b)\\
\hspace{2.3cm}\includegraphics[width=0.425\textwidth,angle=0]{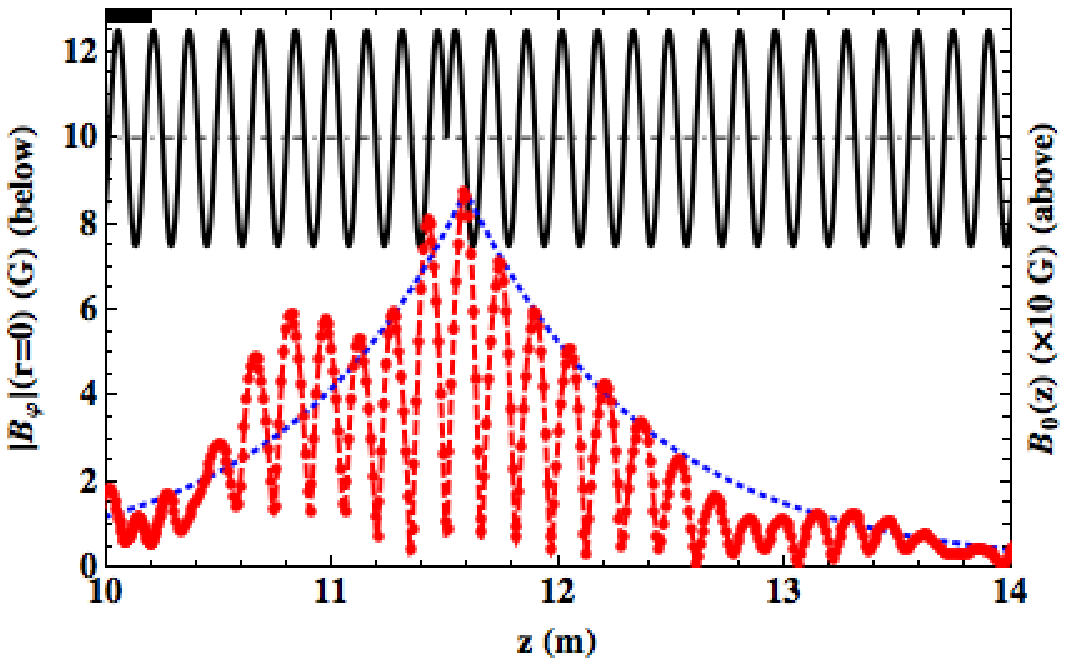}&\hspace{-0.3cm}\includegraphics[width=0.416\textwidth,height=0.263\textwidth,angle=0]{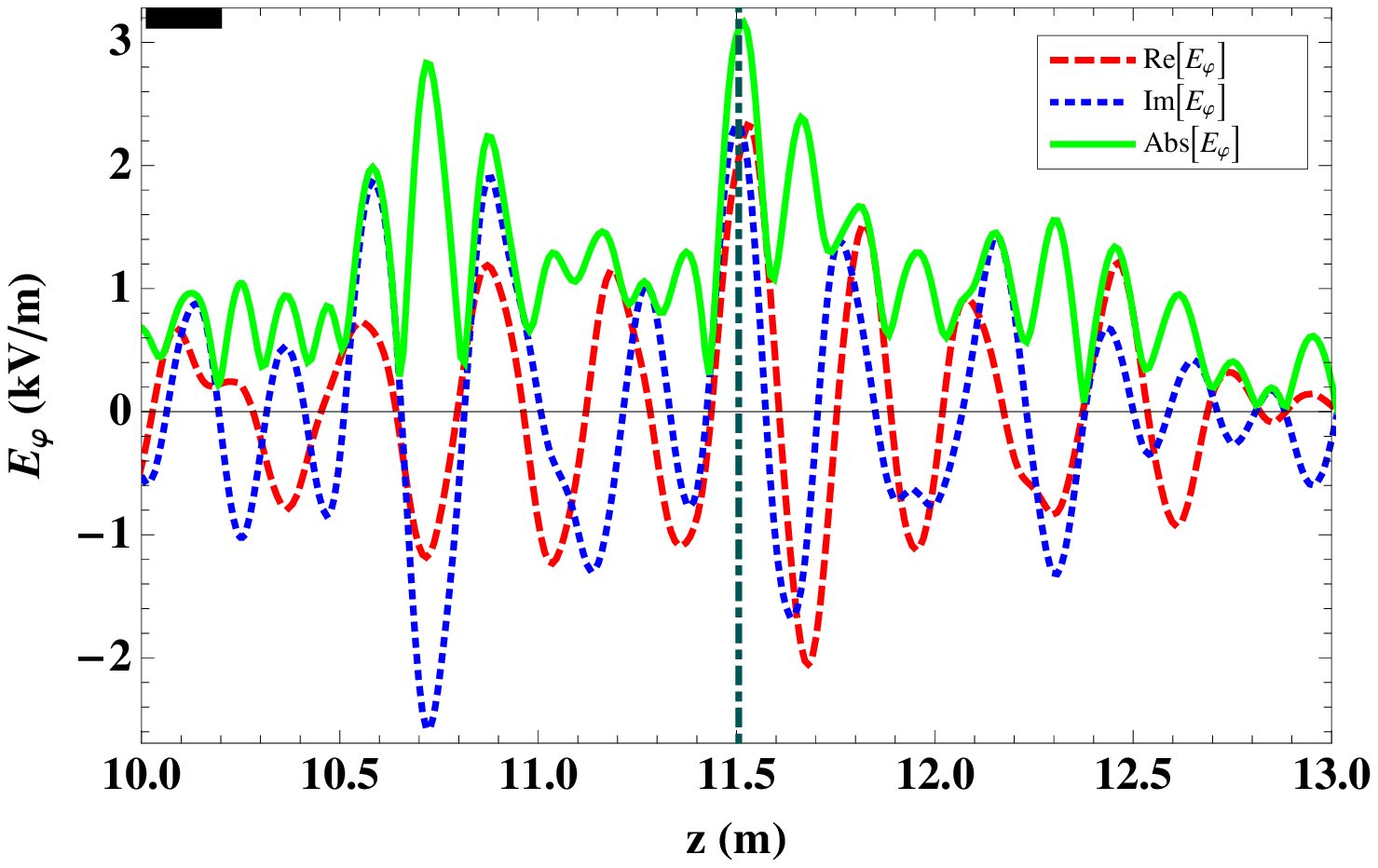}\\
\hspace{2.3cm}(c)&(d)\\
\hspace{2.4cm}\includegraphics[width=0.4\textwidth,height=0.27\textwidth,angle=0]{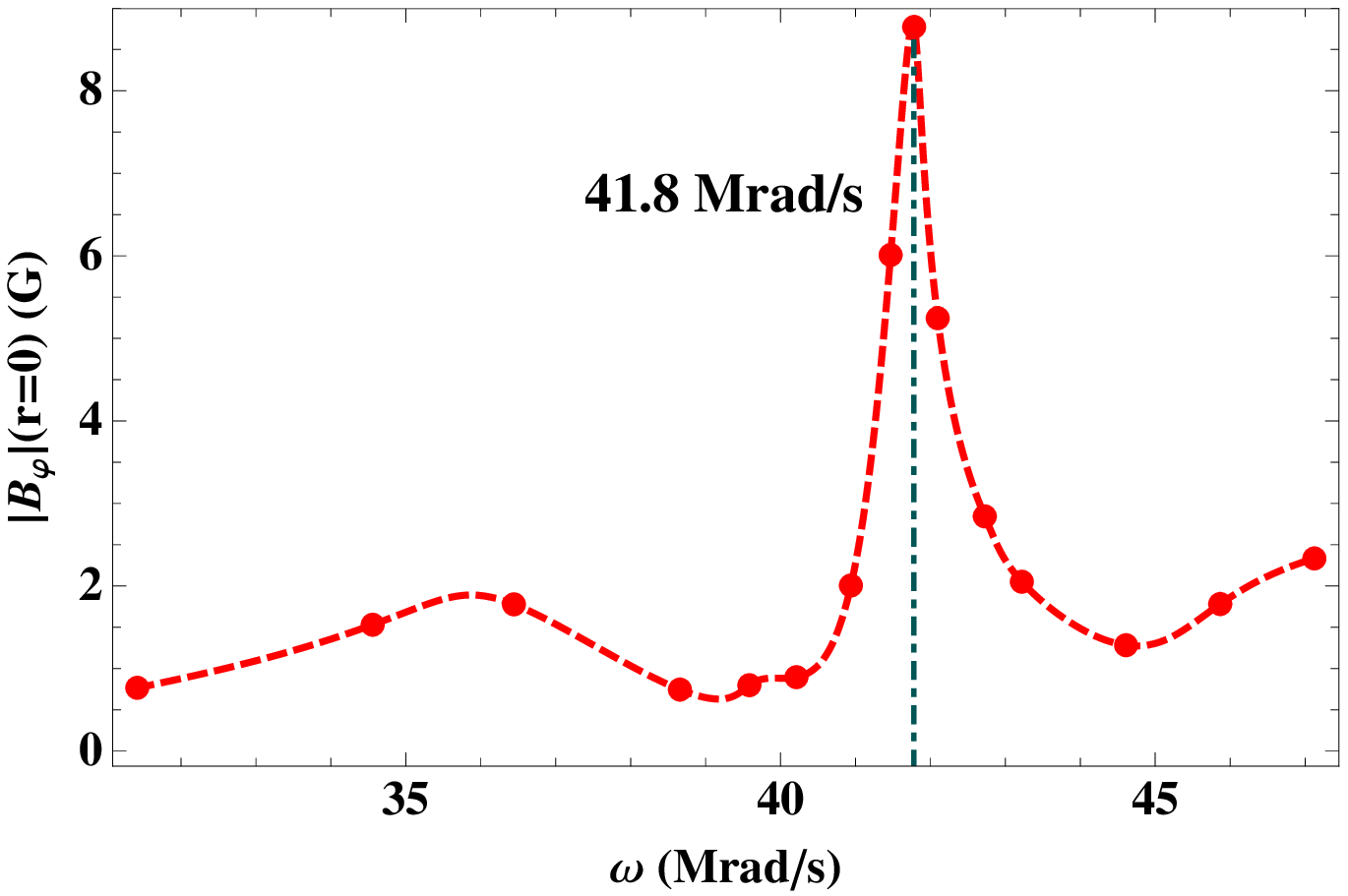}&\hspace{-0.4cm}\includegraphics[width=0.42\textwidth,angle=0]{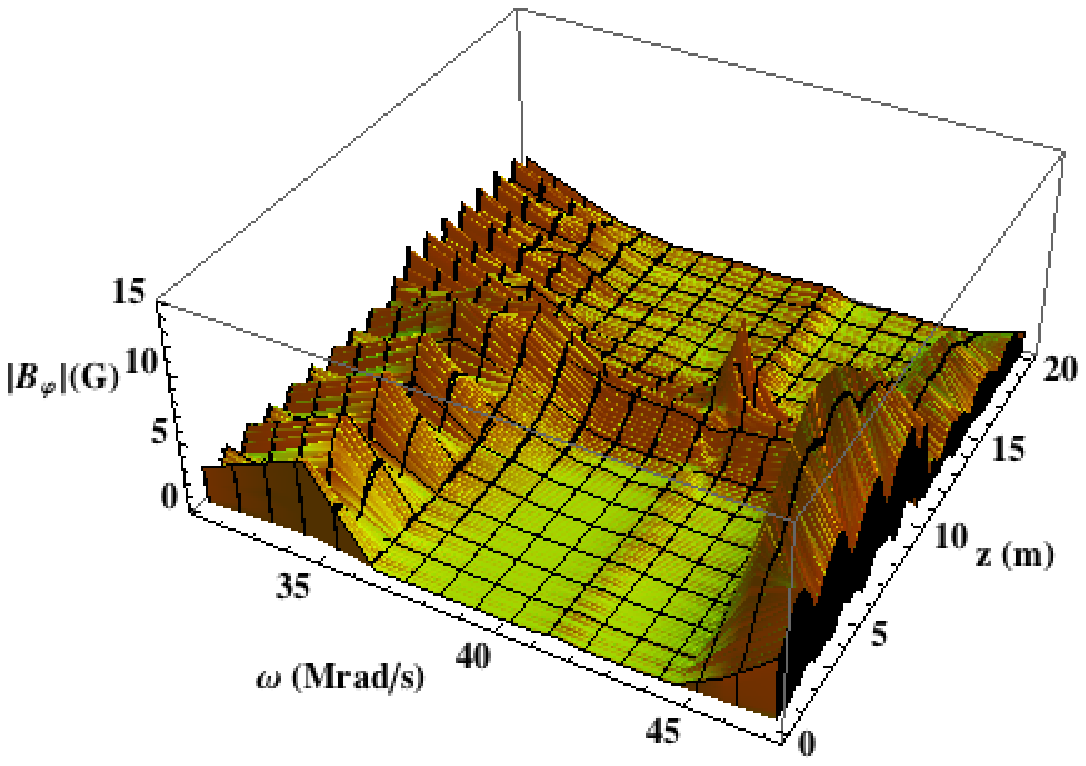}\\
\end{array}$
\end{center}
\caption{Even-parity gap eigenmode: (a) longitudinal profiles of the static magnetic field (solid line) and the rf  magnetic field on axis for $\omega=41.8$~Mrad/s (dots), together with the theoretically calculated envelope (dotted line), (b) longitudinal profile of $E_\varphi$ (on-axis) at $\omega=41.8$~Mrad/s (vertical dot-dashed line marks the defect location, and solid horizontal bar marks the antenna), (c) resonance in the dependence of the on-axis amplitude of the rf magnetic field on driving frequency near the location of the defect, (d) 3D plot of the on-axis wave field strength as a function of $z$ and $\omega$. }
\label{evenmode}
\end{figure}

The effects of collisionality on gap eigenmodes are shown in figure 7, from which we could see clearly a strength drop as the collisionality is increased. 
\begin{figure}[ht]
\begin{center}$
\begin{array}{ll}
\hspace{2.3cm}(a)&(b)\\
\hspace{2.3cm}\includegraphics[width=0.42\textwidth,angle=0]{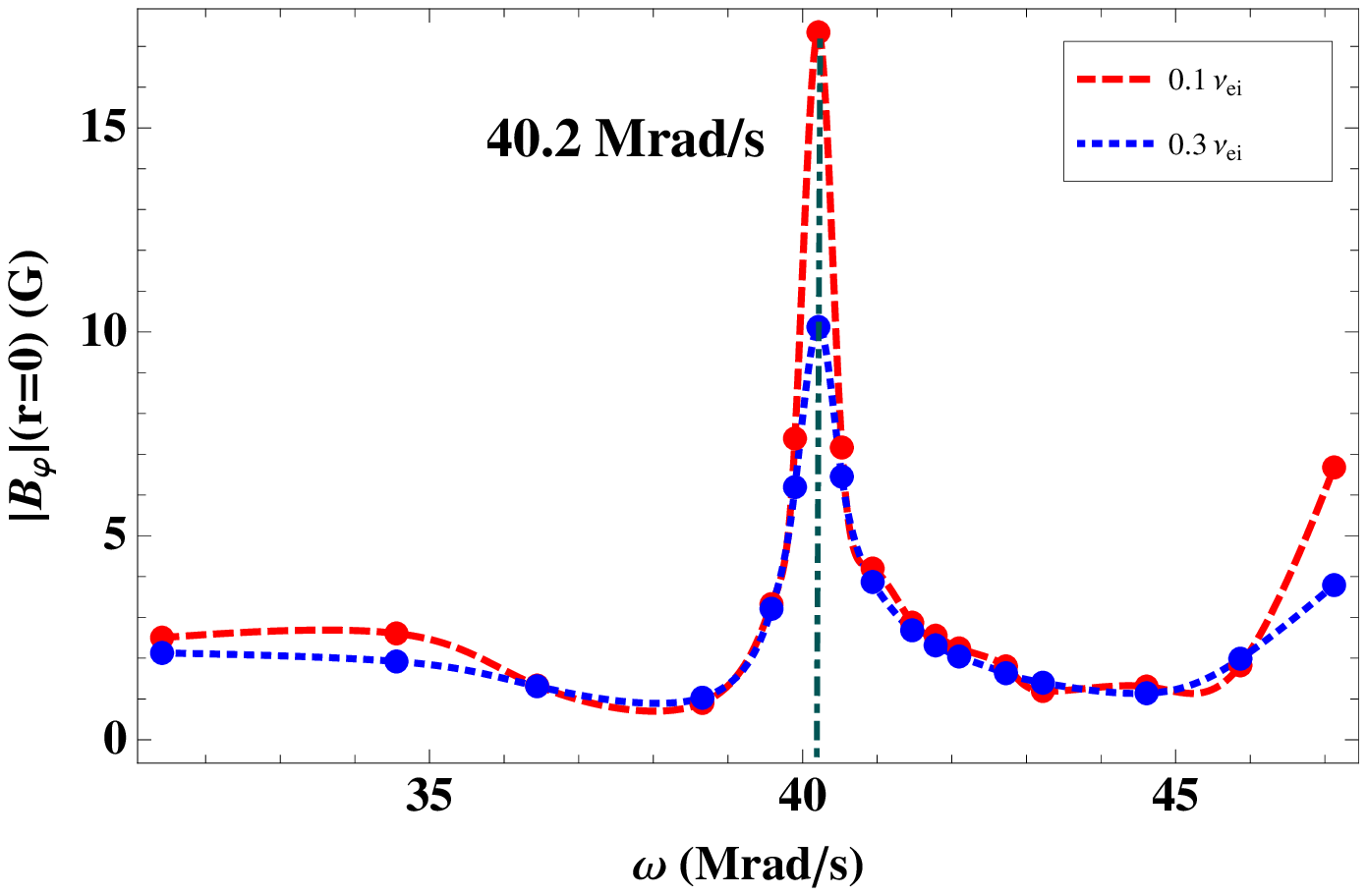}&\includegraphics[width=0.42\textwidth,angle=0]{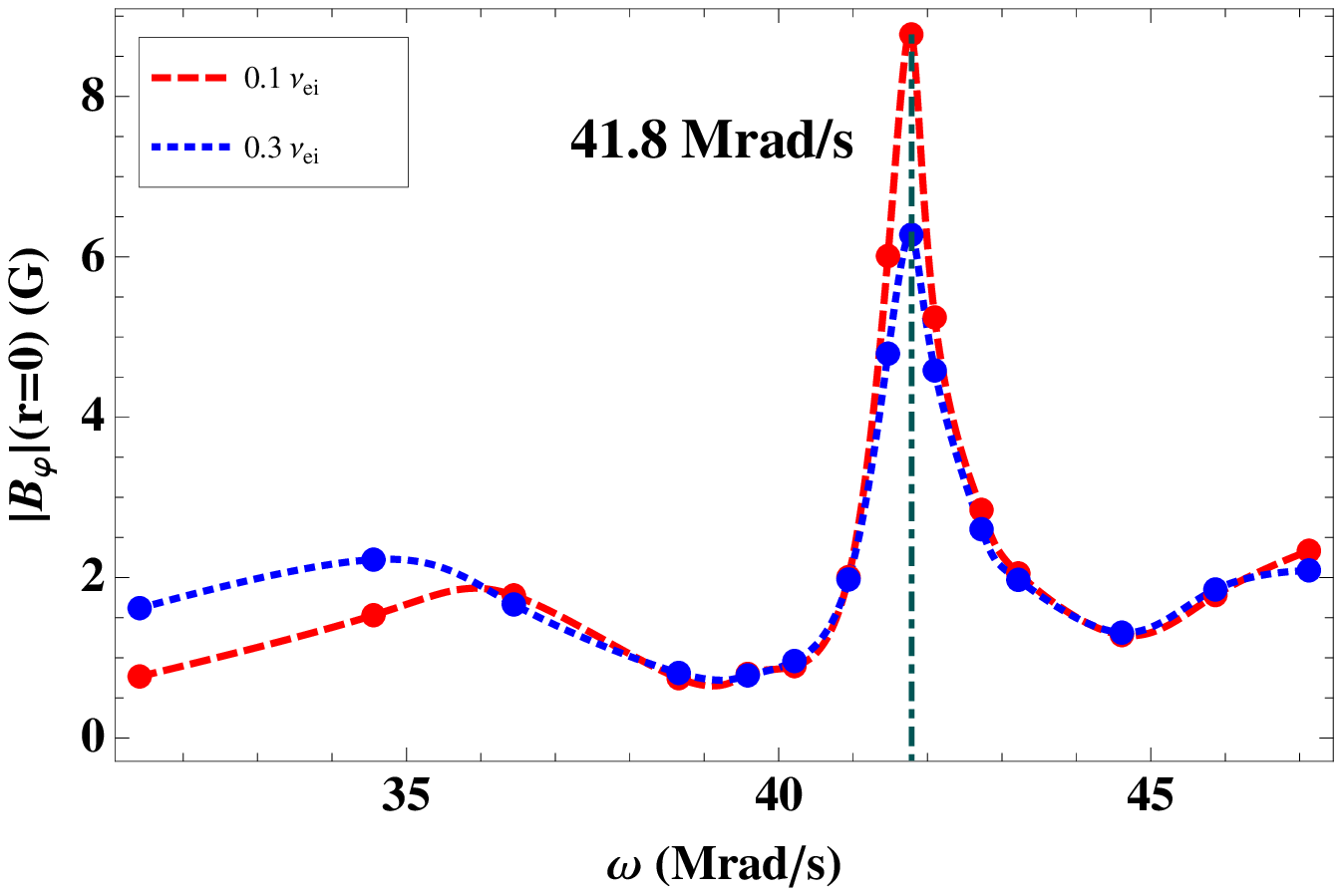}\\
\end{array}$
\end{center}
\caption{Effects of collisionality on gap eigenmodes: (a) results for odd-parity mode (figure 5(c)), (b) results for even-parity mode (figure 6(c)).}
\label{peakwidth}
\end{figure}

\section{Summary}
We have shown that longitudinal modulation of the guiding magnetic field in a plasma column creates a spectral gap for radially localised helicon waves. Calculations performed with an EMS code reveal that this gap prohibits wave propagation along the column when the driving frequency of the rf antenna is in the forbidden range. The calculated width of the gap is consistent with analytical estimates. We have also shown that a discrete eigenmode can been formed inside the spectral gap by introducing a local defect to the periodic structure. Both the theoretical analysis and simulations demonstrate two types of gap eigenmode in the ``imperfect" system: odd-parity mode and even-parity mode, depending the type of the defect employed. The gap mode is localised around the defect and represents a standing wave rather than travelling wave. Its distinctive feature is a resonant peak in the plasma response to the antenna current. The gap eigenmode has two characteristic spatial scales: a short inner scale and a smooth envelope. The inner scale is nearly twice the system's periodicity, which is characteristic for Bragg's reflection; the envelope depends on the modulation amplitude and it scales roughly as the inverse width of the spectral gap. A plausible way to identify the gap mode in a linear device with multi-mirror configuration would be to use the end-plate of the machine as a controllable defect in the periodic system. This could make the mode observable with a modest number of mirrors in the machine. LAPD is an apparent candidate for such experiments, provided that dissipative processes in the plasma do not destroy the gap-mode resonance.

\ack
We would like to thank Dr. Guangye Chen for providing the EMS code and many instructions for its usage. This work was supported by the Chinese Scholarship Council [through scholarchip 2009611029], by the Postgraduate Research Award from the Australian Institute of Nuclear Science and Engineering, by the US Department of Energy under Contract No DE-FG02-04ER54742, and by the Australian Research Council through fellowship FT0991899.

\end{document}